\documentclass[aps,prl,reprint,superscriptaddress,floatfix,showpacs]{revtex4-1}
\usepackage{color}
\usepackage{natbib}
\usepackage{graphicx}
\usepackage{amsmath}
\usepackage{amssymb}
\usepackage{layouts}
\usepackage{tabularx}
\usepackage{array}
\usepackage{textcomp}
\usepackage{docs}
\usepackage{bm}

\usepackage[english]{babel}

\newcommand{\comment}[1]{ }

\begin{document}

\title{Observation of Polarised Microwave Emission from Cosmic Ray Air Showers}

\date{\today}

\author{R.~\v{S}m\'{\i}da}
\email[Corresponding author: ]{radomir.smida@kit.edu}
\affiliation{Karlsruhe Institute of Technology (KIT), Karlsruhe, Germany}
\author{F.~Werner}
\affiliation{Karlsruhe Institute of Technology (KIT), Karlsruhe, Germany}
\author{R.~Engel}
\affiliation{Karlsruhe Institute of Technology (KIT), Karlsruhe, Germany}
\author{J.C.~Arteaga-Vel\'azquez}
\affiliation{Universidad Michoacana, Instituto de F\'{\i}sica y Matem\'aticas, Morelia, M\'{e}xico}
\author{K.~Bekk}
\affiliation{Karlsruhe Institute of Technology (KIT), Karlsruhe, Germany}
\author{M.~Bertaina}
\affiliation{Universit\`{a} di Torino and Sezione INFN, Torino, Italy}
\author{J.~Bl\"{u}mer}
\affiliation{Karlsruhe Institute of Technology (KIT), Karlsruhe, Germany}
\author{H.~Bozdog}
\affiliation{Karlsruhe Institute of Technology (KIT), Karlsruhe, Germany}
\author{I.M.~Brancus}
\affiliation{National Institute of Physics and Nuclear Engineering, Bucharest, Romania}
\author{A.~Chiavassa}
\affiliation{Universit\`{a} di Torino and Sezione INFN, Torino, Italy}
\author{F.~Cossavella}
\altaffiliation[Present address: ]{DLR, Oberpfaffenhofen, Germany}
\affiliation{Karlsruhe Institute of Technology (KIT), Karlsruhe, Germany}
\author{F.~Di~Pierro}
\affiliation{Universit\`{a} di Torino and Sezione INFN, Torino, Italy}
\author{P.~Doll}
\affiliation{Karlsruhe Institute of Technology (KIT), Karlsruhe, Germany}
\author{B.~Fuchs}
\affiliation{Karlsruhe Institute of Technology (KIT), Karlsruhe, Germany}
\author{D.~Fuhrmann}
\altaffiliation[Present address: ]{University of Duisburg-Essen, Duisburg, Germany}
\affiliation{Bergische Universit\"{a}t Wuppertal, Wuppertal, Germany}
\author{C.~Grupen}
\affiliation{Department of Physics, Siegen University, Germany}
\author{A.~Haungs}
\affiliation{Karlsruhe Institute of Technology (KIT), Karlsruhe, Germany}
\author{D.~Heck}
\affiliation{Karlsruhe Institute of Technology (KIT), Karlsruhe, Germany}
\author{J.R.~H\"orandel}
\affiliation{Department of Astrophysics, Radboud University Nijmegen, The Netherlands}
\author{D.~Huber}
\affiliation{Karlsruhe Institute of Technology (KIT), Karlsruhe, Germany}
\author{T.~Huege}
\affiliation{Karlsruhe Institute of Technology (KIT), Karlsruhe, Germany}
\author{K.-H.~Kampert}
\affiliation{Bergische Universit\"{a}t Wuppertal, Wuppertal, Germany}
\author{D.~Kang}
\affiliation{Karlsruhe Institute of Technology (KIT), Karlsruhe, Germany}
\author{H.~Klages}
\affiliation{Karlsruhe Institute of Technology (KIT), Karlsruhe, Germany}
\author{M.~Kleifges}
\affiliation{Karlsruhe Institute of Technology (KIT), Karlsruhe, Germany}
\author{O.~Kr\"{o}mer}
\affiliation{Karlsruhe Institute of Technology (KIT), Karlsruhe, Germany}
\author{K.~Link}
\affiliation{Karlsruhe Institute of Technology (KIT), Karlsruhe, Germany}
\author{P.~{\L}uczak}
\affiliation{National Centre for Nuclear Research, Department of Cosmic Ray Physics, {\L}\'{o}d\'{z}, Poland}
\author{M.~Ludwig}
\affiliation{Karlsruhe Institute of Technology (KIT), Karlsruhe, Germany}
\author{H.J.~Mathes}
\affiliation{Karlsruhe Institute of Technology (KIT), Karlsruhe, Germany}
\author{H.J.~Mayer}
\affiliation{Karlsruhe Institute of Technology (KIT), Karlsruhe, Germany}
\author{S.~Mathys}
\affiliation{Bergische Universit\"{a}t Wuppertal, Wuppertal, Germany}
\author{M.~Melissas}
\affiliation{Karlsruhe Institute of Technology (KIT), Karlsruhe, Germany}
\author{C.~Morello}
\affiliation{Osservatorio Astrofisico di Torino, INAF Torino, Italy}
\author{P.~Neunteufel}
\affiliation{Karlsruhe Institute of Technology (KIT), Karlsruhe, Germany}
\author{J.~Oehlschl\"ager}
\affiliation{Karlsruhe Institute of Technology (KIT), Karlsruhe, Germany}
\author{N.~Palmieri}
\affiliation{Karlsruhe Institute of Technology (KIT), Karlsruhe, Germany}
\author{J.~Pekala}
\affiliation{Institute of Nuclear Physics PAN, Krakow, Poland}
\author{T.~Pierog}
\affiliation{Karlsruhe Institute of Technology (KIT), Karlsruhe, Germany}
\author{J.~Rautenberg}
\affiliation{Bergische Universit\"{a}t Wuppertal, Wuppertal, Germany}
\author{H.~Rebel}
\affiliation{Karlsruhe Institute of Technology (KIT), Karlsruhe, Germany}
\author{M.~Riegel}
\affiliation{Karlsruhe Institute of Technology (KIT), Karlsruhe, Germany}
\author{M.~Roth}
\affiliation{Karlsruhe Institute of Technology (KIT), Karlsruhe, Germany}
\author{F.~Salamida}
\altaffiliation[Present address: ]{Institut Physique Nucl\'{e}aire d'Orsay, Orsay, France}
\affiliation{Karlsruhe Institute of Technology (KIT), Karlsruhe, Germany}
\author{H.~Schieler}
\affiliation{Karlsruhe Institute of Technology (KIT), Karlsruhe, Germany}
\author{S.~Schoo}
\affiliation{Karlsruhe Institute of Technology (KIT), Karlsruhe, Germany}
\author{F.G.~Schr\"oder}
\affiliation{Karlsruhe Institute of Technology (KIT), Karlsruhe, Germany}
\author{O.~Sima}
\affiliation{Department of Physics, University of Bucharest, Bucharest, Romania}
\author{J.~Stasielak}
\affiliation{Institute of Nuclear Physics PAN, Krakow, Poland}
\author{G.~Toma}
\affiliation{National Institute of Physics and Nuclear Engineering, Bucharest, Romania}
\author{G.C.~Trinchero}
\affiliation{Osservatorio Astrofisico di Torino, INAF Torino, Italy}
\author{M.~Unger}
\affiliation{Karlsruhe Institute of Technology (KIT), Karlsruhe, Germany}
\author{M.~Weber}
\affiliation{Karlsruhe Institute of Technology (KIT), Karlsruhe, Germany}
\author{A.~Weindl}
\affiliation{Karlsruhe Institute of Technology (KIT), Karlsruhe, Germany}
\author{H.~Wilczy\'{n}ski}
\affiliation{Institute of Nuclear Physics PAN, Krakow, Poland}
\author{M.~Will}
\affiliation{Karlsruhe Institute of Technology (KIT), Karlsruhe, Germany}
\author{J.~Wochele}
\affiliation{Karlsruhe Institute of Technology (KIT), Karlsruhe, Germany}
\author{J.~Zabierowski}
\affiliation{National Centre for Nuclear Research, Department of Cosmic Ray Physics, {\L}\'{o}d\'{z}, Poland}

\begin{abstract}
We report on the first direct measurement of the basic features of
microwave radio emission from extensive air showers. Using a trigger
provided by the KASCADE-Grande air shower array, the signals of the
microwave antennas of the CROME (Cosmic-Ray Observation via Microwave
Emission) experiment have been read out and searched for signatures of
radio emission by high-energy air showers. Microwave signals have been
detected for more than 30 showers with energies above
$3\times10^{16}$\,eV. The observations presented in this Letter are consistent
with a mainly forward-beamed, coherent and polarised emission process in the GHz
frequency range. An isotropic, unpolarised radiation is disfavoured as the
dominant emission model. The measurements show that microwave radiation offers
a new means of studying air showers at very high energy.
\end{abstract}


\maketitle


\textit{Introduction} -- Since the pioneering studies in the late
1960s~\cite{Allan:1971aa} it has been known that extensive air showers
produce an electromagnetic pulse in the kHz and MHz frequency range in
the Earth's atmosphere. With the availability of much more powerful
electronics and improved shower simulation methods the study of radio
signal emission of air showers has been receiving increasing attention in
recent years. Several new experiments have been set up to study the
characteristics of the radio
signals~\cite{Falcke:2005tc,Ardouin:2006nb,Ardouin:2010gz,Abreu:2012pi}
and different theoretical approaches were developed to understand the
origin of this
radiation~\cite{deVries:2011pa,AlvarezMuniz:2011bs,Huege:2013vt}.

Extensive air showers consist of a disk of high-energy particles
traversing the atmosphere. With the thickness of this disk being of
the order of $1$\,m, charged particles are expected to emit
electromagnetic waves coherently at frequencies below about
$100$\,MHz. It is now understood that the observed radio signal stems
from several emission processes~\cite{Huege:2010vm}. First of all the
electrons and positrons of the shower disk are deflected in the
Earth's magnetic field (geomagnetic
radiation)~\cite{Huege:2003up}. Secondly, there are about
$20$--$30$\,\% more electrons than positrons in the shower disk
leading to a varying charge excess and, hence, electromagnetic
radiation (Askaryan effect)~\cite{Askaryan:1961x1,Askaryan:1965x1,Kahn:1966x1}.

Until recently, radio emission from air showers in the GHz range has not been
considered a promising detection channel as the typical
wavelengths are in the cm range. Correspondingly very few measurements
exist. Early attempts covering frequencies up to $550$\,MHz
confirmed the strong, exponential suppression of higher
frequencies~\cite{Fegan:1968aa,Spencer:1969x1,Fegan:1969x1,Fegan:2011fb}.
In one of the balloon flights of ANITA (Antarctic Impulsive
Transient Antenna), radio signals of extensive air showers were
discovered serendipitously and found to extend even up to
$900$\,MHz~\cite{Hoover:2010qt} (see also~\cite{AlvarezMuniz:2012sa}).

In addition to the aforementioned emission processes, a third source
of GHz radiation could be present in air showers. Gorham et
al.~\cite{Gorham:2007af} pointed out that the numerous slow electrons from ionisation
produced by the high-energy particles of the shower disk are
expected to emit molecular bremsstrahlung at GHz frequencies.  Several
experiments were set up to search for such a signal with test
beams~\cite{Gorham:2007af,Monasor:2011ft,Alvarez-Muniz:2013x1,Bluemer:2013aa}
and air shower
detectors~\cite{Gorham:2007af,AlvarezMuniz:2012ew,Facal:2013x1,Smida:2011cv}. First
unambiguous detections of microwave signals of air showers in the
$3$--$4$\,GHz range were reported recently by two of these
experiments, EASIER at the Pierre Auger Observatory~\cite{Facal:2013x1}
and CROME~\cite{Smida:2013x1}.

In this work we present the first study of the basic features of the
microwave radiation of air showers using data of the CROME
experiment. We show that the emission takes place mainly within the
Cherenkov ring around the shower axis.  Comparing the CROME data with
theoretical predictions we conclude that the observations are
compatible with a superposition of geomagnetic and Askaryan emission
processes. Possible applications of the GHz radiation include the
measurement of extensive air showers at very high energy with a duty
cycle of virtually $100$\,\% as compared to typically $15$\,\% for
experiments using Cherenkov light.


\textit{The CROME Experiment} -- The CROME
experiment~\cite{Smida:2011cv} is located within the KASCADE-Grande
air shower array~\cite{Apel:2010zz} at the Karlsruhe Institute of
Technology. It consists of different radio antennas covering a wide
range of frequencies. The readout of the antennas is triggered by
KASCADE-Grande, which is optimised for the detection of air showers in
the range from $3\times10^{15}$ to $10^{18}$\,eV. The trigger signal
for CROME is built on a coincident measurement of the three innermost clusters
of KASCADE-Grande stations.

The data reported in this article were taken with antennas measuring
in the extended C~band ($3.4$--$4.2$\,GHz) range. These C~band
detectors consist of a parabolic reflector of $335$\,cm diameter
($\sim40$\,dBi gain) and a multi-receiver camera. In total three
antennas of this type were installed and operated over different
periods of time between May 2011 and November 2012. One antenna was
directed vertically upward and the other two were tilted relative to
the first one by $\pm 15$\textdegree\ with respect to magnetic North.

The cameras were equipped with $9$ linearly polarised C~band receivers
consisting of reflector-matched feed horns and low noise blocks (LNBs,
Norsat 8215F). The receivers were arranged in a compact $3\times3$
matrix in the focal plane of the reflector. The four corner feed horns
were equipped with a second LNB in each camera during summer 2012,
allowing the measurement of two polarisation directions for the same
signal direction.

The radiation pattern of the antenna system was measured with a
calibrated airborne transmitter~\cite{Smida:2011cv}. The difference
between the main and side lobe of the receivers is more than $10$\,dB
even in off-center channels which are affected by aberration
effects. The half-power beam width is less than $2$\textdegree\ for
all channels. The estimated system noise temperature is about $50$\,K.

A logarithmic power detector (amplifier) was used to measure the envelope
of the antenna signals within an effective bandwidth of $\sim 600$\,MHz.
The response time of the complete system is $\sim4$\,ns (exponential time
constant), allowing the measurement of even very short pulses. The
signal is digitised and stored for $10$\,\textmu s before and after
the trigger delivered by KASCADE-Grande.

The reconstructed arrival direction, core position, and energy of the
showers recorded by the KASCADE-Grande array are used in the further
analysis. The reconstruction accuracy is about $0.8$\textdegree\ for the
arrival direction, approximately $6$\,m for the core position and about
$20$\,\% for the energy if the standard quality cuts are
applied~\cite{Apel:2010zz}.


\begin{figure}[htb!]
\begin{center}
\includegraphics{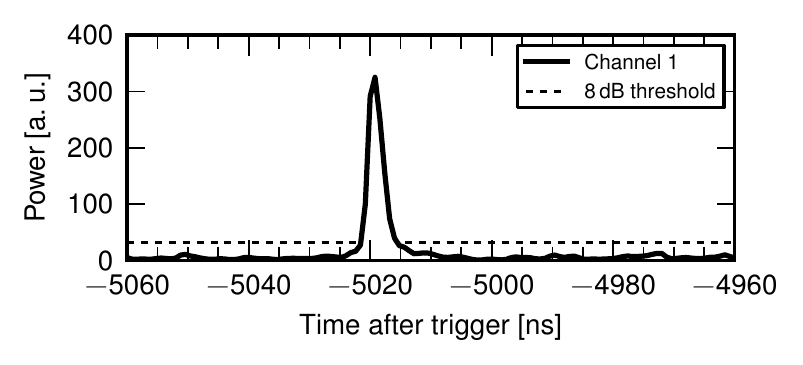}\\
\includegraphics{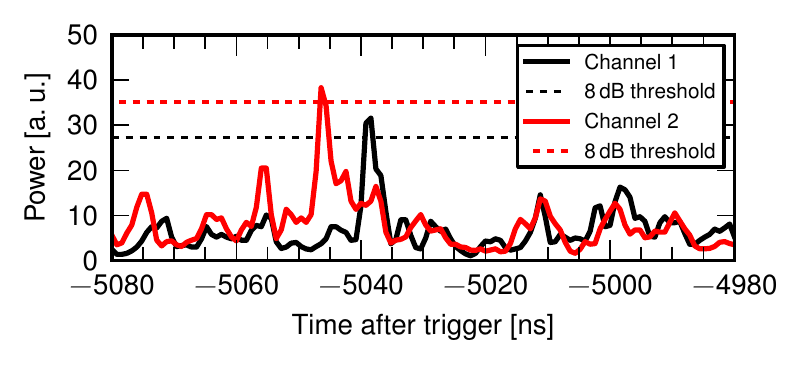}
\caption{Power received by the logarithmic amplifier as function of
  time relative to the KASCADE-Grande trigger. Shown are the event
  with the highest signal (top) and a stereo event (bottom). The
  signal thresholds of $8$\,dB are shown as dashed lines.
\label{Fig-event}
}
\end{center}
\end{figure}

\textit{Event selection} -- We have analysed all reconstructed events
passing the KASCADE-Grande quality cuts
measured up to August 10, 2012.
From these events, we selected the showers crossing the field of view
of at least one CROME receiver (approximately $5.5$ showers above
$3\times10^{16}$\,eV per day) for which a cone of half-width
of $2$\textdegree\ was assumed.

The expected arrival time of the microwave signal from each air shower
was calculated using the reconstructed air shower geometry, accounting
for the altitude-dependent refractive index and the measured signal
propagation times in the detectors. The typical uncertainty of the
signal arrival time due to measurement, reconstruction, and simulation
uncertainties is about $50$\,ns. The strength of a signal in the
estimated time window is quantified relative to the mean noise level
of each $20$\,\textmu s trace.

Selecting events with a pulse amplitude of more than $8$\,dB above the
mean trace voltage and energy above $3\times10^{16}$\,eV we have found
$35$ showers with a microwave signal. Two of those showers produced a
signal above the threshold level in two microwave receivers with the
predicted time ordering. The expected number of noise signals above
$8$\,dB is estimated from data by repeating the analysis for shifted time
windows and it is found to be $7.1\pm1.6$.

A signal with an amplitude of more than $10$\,dB, corresponding to
more than $5\,\sigma$ significance for the observed baseline
fluctuations within a $50$\,ns time window, was measured for $8$
events. For the given time interval no noise signal of the same
strength is expected.

The time traces of microwave signals measured for two air showers are
shown in Fig.~\ref{Fig-event}. The top panel shows the largest signal,
$17.7$\,dB above the noise level, measured with the CROME
experiment. The energy of the shower was $2.5\times10^{17}$\,eV with a
zenith angle of $5.6$\textdegree\ and a core distance of $120$\,m to the
antenna. A shower measured in two receivers of two different antennas
(stereo event) is shown in the bottom panel. The shower parameters are:
$3.7\times10^{16}$\,eV, zenith angle $3.7$\textdegree\ and core
distance $110$\,m. Very good agreement is found between the measured
and calculated time differences for the signals in the two independent
antennas in the case of the stereo event.


\textit{Properties of the microwave signal} -- The duration of the
microwave signals is about $10$\,ns. Based on the shower geometry and
field of view of the corresponding receivers it can be concluded that
most of the signals were emitted from altitudes close to the expected
maximum of shower development (typically $4$\,km above the ground).

\begin{figure}[htb!]
\begin{center}
\includegraphics{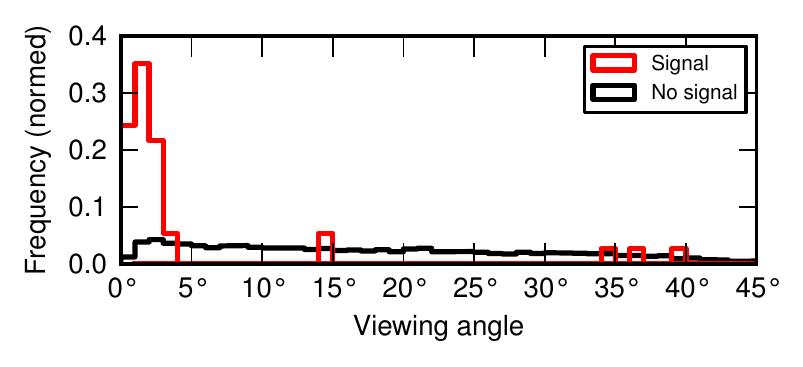}
\caption{Distribution of the viewing angle for showers passing the
  KASCADE-Grande and geometry selection criteria.  The red histogram
  shows the events with a microwave signal greater than $8$\,dB. The
  black histogram shows all other events.
\label{Fig-viewing-angles}
}
\end{center}
\end{figure}

In Fig.~\ref{Fig-viewing-angles}, the distribution of the viewing
angle (the angle between the shower axis and the boresight of the
receiver) is shown for events passing the shower selection
criteria. The red line indicates the distribution of the viewing angle
for receivers with a signal above $8$\,dB and the black line for all
other receivers. One can clearly notice a lack of events in the signal
distribution at viewing angles larger than $4$\textdegree. With a
receiver field of view of $\sim2$\textdegree\ the observed angles
are compatible with the Cherenkov angle in air ($\sim1$\textdegree).

\begin{figure}[t]
\begin{center}
\includegraphics{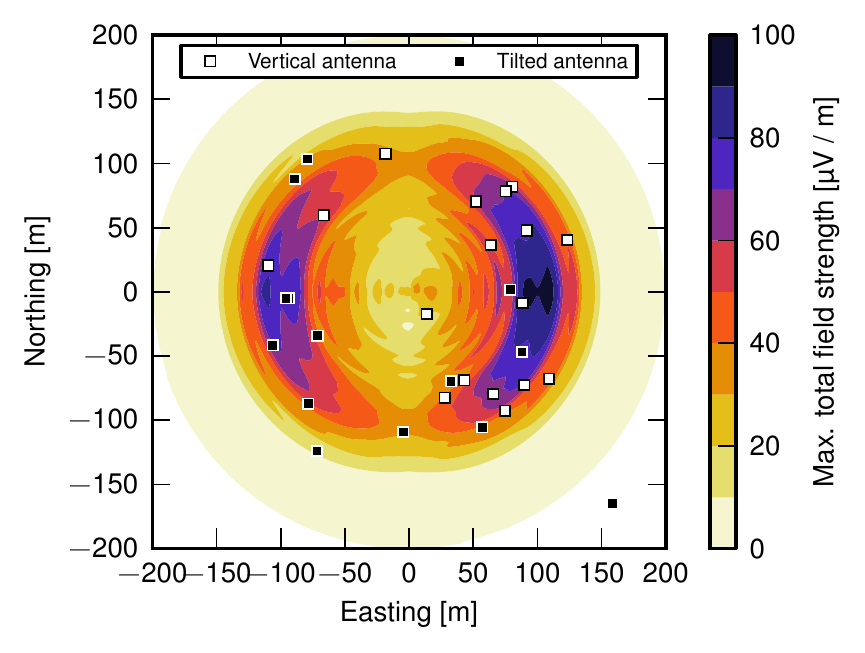}\\
\includegraphics{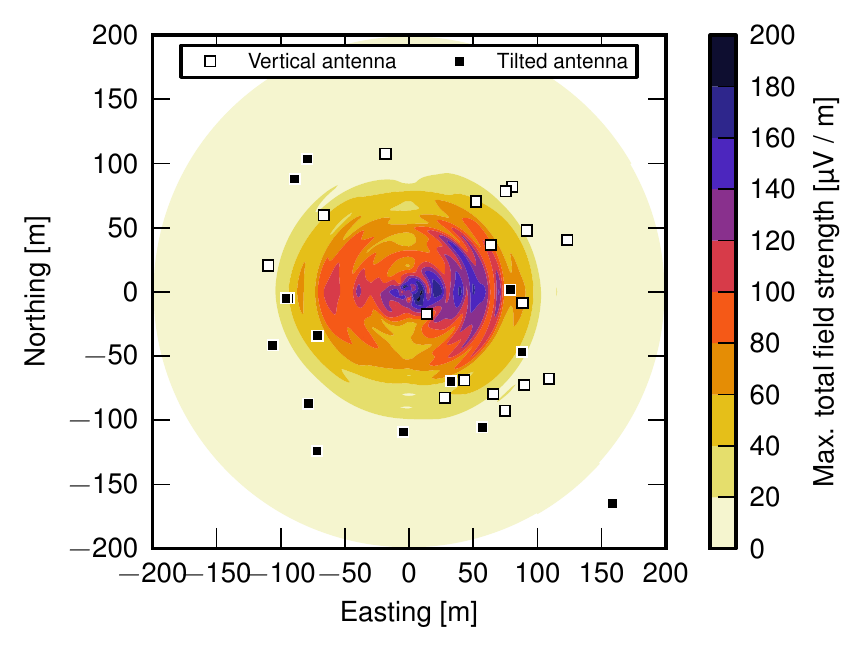}
 \caption{Positions at which a microwave signal has been detected
   relative to the shower core at $(0,0)$. The maximum total field
   strength predicted by CoREAS for a typical vertical
   (upper plot) and a very deep vertical proton shower (lower panel)
   of $10^{17}$\,eV is shown.
   \label{Fig-cores}
 }
\end{center}
\end{figure}

In addition, the core position of events with microwave signal form
a ring structure at a distance of $70$--$150$\,m around the antennas
(cf.~Fig.~\ref{Fig-cores}).
All these features favour an emission mechanism with properties
similar to Cherenkov radiation as the dominant source of the observed
GHz signals.


\textit{Interpretation} -- For a comparison of the measured events
with predictions for the radio signal expected due to the geomagnetic and
Askaryan emission mechanisms we improve the purity of the event
sample. By considering only events with a viewing angle less than
$4$\textdegree\ (cf.~Fig.~\ref{Fig-viewing-angles}) we obtain $30$ showers
(two of which were observed
with two receivers) for an expected number of $1.2\pm0.5$ noise signals.
For these events, the positions at which the GHz signal is detected
relative to the shower core are given in Fig.~\ref{Fig-cores}.

By applying the end-point formalism~\cite{James:2010vm},
CoREAS~\cite{Huege:2013vt} allows the simulation of the radio emission associated with
acceleration of charged particles in extensive air showers
(i.e.\ including signals due to both the geomagnetic and Askaryan
effects).
At GHz frequencies, the predicted electromagnetic pulses form a
Cherenkov-like cone in the forward direction.
The electric field has its highest amplitude on a ring at a distance
of $100\pm20$\,m. The radius of this ring structure is directly correlated
with the depth of shower maximum~\cite{AlvarezMuniz:2012sa,
Huege:2013vt,deVries:2013dia}.

The expected electric field strength of a radio pulse calculated with
the CoREAS simulation package for a vertical shower
is shown in Fig.~\ref{Fig-cores} to illustrate basic features of
the expected radio signal at ground. The predicted electric field
is shown for two vertical showers of $10^{17}$\,eV:
a typical shower with the depth of maximum at
$X_{\textnormal{max}}=658$\,g\,cm$^{-2}$ (upper panel) and a deep proton
shower with the shower maximum at $X_{\textnormal{max}}=895$\,g\,cm$^{-2}$
(lower panel).
Both the structure of the Cherenkov-like ring and the asymmetries observed in data
are qualitatively well reproduced by the typical shower.

In CoREAS, the major emission mechanism in the GHz range is the
geomagnetic process with a small contribution coming from the time
varying charge excess (Askaryan process). Considering nearly vertical
showers, the superposition of the mainly east-west polarised electric
field of geomagnetic radiation with the radially inward polarised
field due to the Askaryan effect leads to a pronounced east-west
asymmetry in the overall signal strength. This asymmetry is confirmed
by the data taken with the vertically upward pointing antenna (open
symbols in Fig.~\ref{Fig-cores}), which has the largest exposure of
all three CROME antennas. There are $14$ and $3$ events detected with
the antenna east and west of the shower core, respectively.

The statistics of events measured with tilted antennas (filled symbols
in Fig.~\ref{Fig-cores}) is dominated by the antenna pointing towards
north.
The angle between this antenna and the local geomagnetic field vector
is about $40$\textdegree.
No significant east-west asymmetry is observed in the core distribution
measured by this antenna which can be expected for an increasing dominance
of geomagnetic emission. The event rate measured by the antenna
pointed almost parallel with the local geomagnetic field vector is much
smaller than in the other two antennas.

A detailed comparison of the observed signal amplitudes with the CoREAS
predictions can only be done after a full end-to-end calibration of CROME
and is beyond the scope of this article.

\begin{figure}[htb!]
\begin{center}
\includegraphics{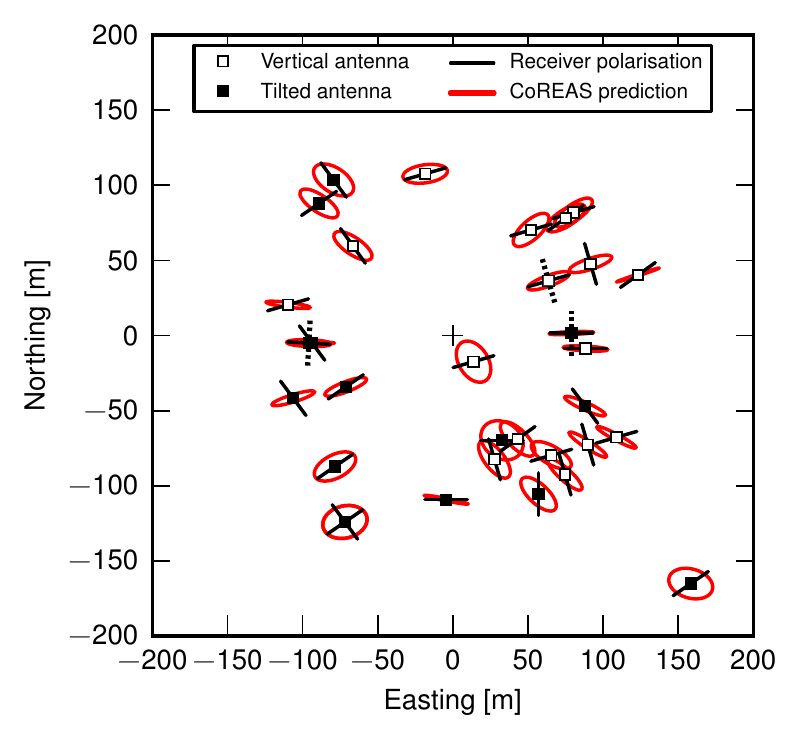}
 \caption{Polarisation directions of receivers in which a microwave
   signal was detected (black lines). In addition, the predicted
   polarisation ellipses simulated with CoREAS for iron showers are
   shown in red. For dual-polarised receivers, the polarisation
   direction in which no signal was detected is shown as dashed
   line. The shower core is always at $(0,0)$.
   \label{Fig-polarisation}
 }
\end{center}
\end{figure}

To compare the measured polarisation directions with CoREAS
predictions we simulated showers for each observed event matching the
geometry, energy, and muon number reconstructed with
KASCADE-Grande. In~Fig.~\ref{Fig-polarisation} the polarisation
directions of the receivers with a microwave signal are shown together
with those obtained from the CoREAS simulations.

The polarisation information can be compared directly with the CoREAS
prediction for the three events detected with dual-polarised
receivers. It is found that the signal was always detected only in the
receiver with the polarisation direction close to that predicted in
the simulations, but the statistics of such events is too small to
draw conclusions.

Therefore we applied a detector simulation to the time traces obtained
from the CoREAS simulations and calculated the loss in detectable
power due to the projection of the predicted electric field vector
onto the polarisation direction of the receivers (polarisation
loss). The polarisation loss must be small particularly for showers
with lower energies.
A large number of events was indeed detected in receivers whose polarisation
direction matches favourably the local electric field vector of the shower.

Assuming that the time-dependent local electric field vector is
correctly described by the CoREAS predictions we find an average
polarisation loss of $38$\,\%. For unpolarised pulses from incoherent
radiation with a flat frequency spectrum, an average polarisation loss
of $50.0\,\% \pm 2.5\,\%$ would be expected. Therefore the comparison
with CoREAS yields a significance of $5\,\sigma$ that the radiation is
not unpolarised. The agreement between the polarisation pattern
predicted by CoREAS and our measurements is also better than that of
other simplified models, e.g.\ linear polarisation along the
east-west, north-south, and radial directions.

Finally, we also studied high-energy events with a favourable geometry
but without a detected microwave signal above $8$\,dB. We found that
the lack of microwave signals could be explained by the large angles
between the polarisation directions of the receivers and those expected
for the corresponding air showers.

\textit{Conclusions and Outlook} -- Using showers measured with CROME
in coincidence with the KASCADE-Grande air shower array we have
determined fundamental properties of the microwave emission of air
showers in the forward direction. The spatial and angular distributions of
the microwave signal resemble those expected for Cherenkov light
emission. The collected polarisation information is incompatible with
the hypothesis of an unpolarised signal at the $5\,\sigma$ level.

A comparison with CoREAS simulations showed that the measurements are
qualitatively in good agreement with the extension of the well-known
radio emission processes at tens of MHz into the GHz range due to the
time compression of the signal close to the Cherenkov
angle~\cite{deVries:2011pa,AlvarezMuniz:2012sa,Huege:2013vt}.
We conclude that the measurements favour a coherent emission process that
is dominated by an emission region close to the depth of shower maximum.
These findings, however, do not exclude a sub-leading
signal component due to an isotropic emission process as expected for
molecular bremsstrahlung.

Benefiting from the low background noise and the nearly perfect
transparency of the atmosphere at microwave frequencies, as well as
the availability of a well-developed technology of microwave
detection~\cite{Gorham:2007af} one can envisage various applications
of measuring air showers with setups similar to the one presented
here.

For example, measurements with setups similar to CROME can serve as
experimental cross-check and calibration of the expected signal for
the balloon-borne ANITA detector at the South
Pole~\cite{Hoover:2010qt}. The forthcoming ANITA measurement campaign
will aim at detecting a large number of ultra-high energy cosmic rays
by observing their microwave signal reflected off the Antarctic ice.

Additionally, compared to optical detectors such as imaging atmospheric
Cherenkov telescopes~\cite{Voelk:2008fw}, the possibility of measuring
with a nearly $100$\,\% duty cycle and using simple metallic
reflectors instead of optical mirrors could make this measurement
technique a promising technology for showers of very high energy above
$10^{15}$\,eV. To reach an angular resolution similar to that of current
$\gamma$-ray telescopes, however, a parabolic reflector of $20$--$30$\,m
diameter would be needed.


\textit{Acknowledgments} -- It is our pleasure to acknowledge the
interaction and collaboration with many colleagues from the Pierre
Auger Collaboration, in particular Peter Gorham, Antoine
Letessier-Selvon and Paolo Privitera. This work has
been supported in part by the KIT start-up grant 2066995641, the
ASPERA project BMBF 05A11VKA and 05A11PXA, the Helmholtz-University
Young Investigators Group VH-NG-413 and the National Centre for
Research and Development in Poland (NCBiR) grant ERA-NET-ASPERA/01/11.


\end{document}